# PSMACA: An Automated Protein Structure Prediction Using MACA (Multiple Attractor Cellular Automata)

P. Kiran Sree[1], Inampudi Ramesh Babu[2], and N. Usha Devi[3, *]

[1]*Department of Computer Science and Engineering, Jawaharlal Nehru Technological University, Hyderabad 500085, India*
[2]*Department of Computer Science and Engineering, Acharya Nagarjuna University, Guntur 522510, India*
[3]*Department of Computer Science and Engineering, Jawaharlal Nehru Technological University, Kakinada 533003, India*

Protein Structure Predication from sequences of amino acid has gained a remarkable attention in recent years. Even though there are some prediction techniques addressing this problem, the approximate accuracy in predicting the protein structure is closely 75%. An automated procedure was evolved with MACA (Multiple Attractor Cellular Automata) for predicting the structure of the protein. Most of the existing approaches are sequential which will classify the input into four major classes and these are designed for similar sequences. PSMACA is designed to identify ten classes from the sequences that share twilight zone similarity and identity with the training sequences. This method also predicts three states (helix, strand, and coil) for the structure. Our comprehensive design considers 10 feature selection methods and 4 classifiers to develop MACA (Multiple Attractor Cellular Automata) based classifiers that are build for each of the ten classes. We have tested the proposed classifier with twilight-zone and 1-high-similarity benchmark datasets with over three dozens of modern competing predictors shows that PSMACA provides the best overall accuracy that ranges between 77% and 88.7% depending on the dataset.

## 1. INTRODUCTION

Proteins are molecules with macro structure that are responsible for a wide range of vital biochemical functions, which includes acting as oxygen, cell signaling, antibody production, nutrient transport and building up muscle fibers. Specifically, the proteins are chains of amino acids, of which there are 20 different types, coupled by peptide bonds.[2] The three-tiered structural hierarchy possessed by proteins is typically referred to as primary and tertiary structure. Protein Structure Predication from sequences of amino acid gives tremendous value to biological community. This is because the higher-level and secondary level[1, 2] structures determine the function of the proteins and consequently, the insight into its function can be inferred from that.

As genome sequencing projects are increasing tremendously. The SWISS-PORT databases[3, 4] of primary protein structures are expanding tremendously. Protein Data Banks are not growing at a faster rate due to innate difficulties in finding the levels of the structures. Structure determination[5, 6] procedure experimental setups will be very expensive, time consuming, require more labor and may not applicable to all the proteins. Keeping in view of shortcomings of laboratory procedures in predicting the structure of protein major research have been dedicated to protein prediction of high level structures using computational techniques. Anfinsen did a pioneering work predicting the protein structure from amino acid sequences.[6, 7] This is usually called as protein folding problem which is the greatest challenge in bioinformatics. This is the ability to predict the higher level structures from the amino acid sequence.

By predicting the structure of protein the topology of the chain can be described. The tree dimensional arrangement of amino acid sequences can be described by tertiary structure. They can be predicted independent of each other. Functionality of the protein can be affected by the tertiary structure, topology and the tertiary structure. Structure aids in the identification of membrane proteins, location of binding sites and identification of homologous proteins[9–11] to list a few of the benefits, and thus highlighting the importance, of knowing this level of structure. This is the reason why considerable efforts have been devoted in predicting the structure only. Knowing the structure of a protein is extremely important and can also greatly enhance the accuracy of tertiary structure prediction. Furthermore, proteins can be classified according to their structural elements, specifically their alpha helix and beta sheet content.

## 2. RELATED WORKS IN STRUCTURE PREDICTION

The Objective of structure prediction is to identify whether the amino acid residue of protein is in helix, strand or any other shape. In 1960 as a initiative step of structure prediction the probability of respective structure element is calculated for each amino acid by taking single amino acid properties consideration.[1,3,6] This method of structure prediction is said to be first generation technique. Later this work extended by considering the local environment of amino acid said to be second generation technique. In case of particular amino acid structure prediction adjacent residues information also needed, it considers the local environment of amino acid it gives 65% structure information. So that extension work gives 60% accuracy. The third generation technique includes machine learning, knowledge about proteins, several algorithms which gives 70% accuracy. Neural networks[10,11] are also useful in implementing structure prediction programs like PHD, SAM-T99.

The evolution process is directed by the popular Genetic Algorithm (GA) with the underlying philosophy of survival of the fittest gene. This GA framework can be adopted to arrive at the desired CA rule structure appropriate to model a physical system. The goals of GA formulation are to enhance the understanding of the ways CA performs computations and to learn how CA may be evolved to perform a specific computational task and to understand how evolution creates complex global behavior in a locally interconnected system of simple cells.

Techniques for structure prediction include, but are not limited to, constraint programming methods, statistical approaches to predict the probability of an amino acid being in one of the structural elements, and Bayesian network models.[12,13] Nearest neighbor techniques attempt to predict the structure of a central residue, within a segment of amino acids, based on the known structures of homologous segments. In, a technique based on multiple linear regressions was presented to predict structure. Published techniques for structure prediction span over a period of three decades, with the early works of Lim and Chou and Fasman in the 1970s.

## 3. CELLULAR AUTOMATA

Cellular Automata (CA) is a simple model of a spatially extended decentralized system, made up of a number of individual components (cells). The communication among constituent cells is limited to local interaction. Each individual cell is in a specific state that changes over time depending on the states of its neighbors. From the days of Von Neumann who first proposed the model of Cellular Automata (CA),[24,25] to Wolfram's recent book 'A New Kind of Science,' the simple and local neighborhood structure of CA has attracted researchers from diverse disciplines. It has been subjected to rigorous mathematical and physical analysis for past fifty years and its application has been proposed in different branches of science—both social and physical.

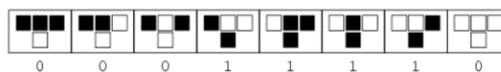

**Fig. 1.** Example of rule formation (Rule 30).

DEFINITION. CA is defined a four tipple $\langle G, Z, N, F \rangle$

Where $G \rightarrow$ Grid (Set of cells)
$Z \rightarrow$ Set of possible cell states
$N \rightarrow$ Set which describe cells neighborhoods
$F \rightarrow$ Transition Function (Rules of automata)

The concept of the homogeneous structure of CA was initiated in early 1950s by Neumann.[20,22] It was conceived as a general framework for modeling complex structures, capable of self-reproduction and self-repair. Subsequent developments have taken place in several phases and in different directions.

Dr. Stephen Wolfram referred to as Rule 30 in Figure 1, produces a binary sequence that is sufficiently random and can be used as a secure encryption system. Rules are formed through a definition of the $2^3 = 8$ possible progressions of three cells (the cell, the cells left-hand neighbor, and the cells right-hand neighbor). Each of these progressions gives a single output, producing a new cell and creating a three to one mapping. The Rules are then named using these progressions as shown in figure. The name of the Rule can be found by arranging the progressions, starting from the left with seven base two $(111)_2$, descending to zero $(000)_2$, and converting this base two number to base ten. In doing this, there are $2^8 = 256$ possibilities, and therefore 256 possible rules. The name of each rule is given by the base 10 representation of their output. This is the set of parameters and outputs for Rule 30.

## 4. DESIGN OF MACA BASED PATTERN CLASSIFIER

An $n$-bit MACA with $k$-attractor basins can be viewed as a natural classifier. It classifies a given set of patterns into $k$ number of distinct classes, each class containing the set of states in the attractor basin. To enhance the classification accuracy of the machine, most of the works have employed MACA as in Figure 2, to classify patterns into two classes (say I and II). The following

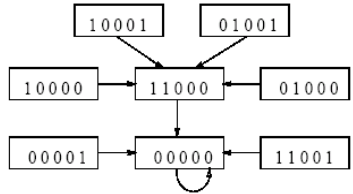

**Fig. 2.** Example of MACA with basin 0000.

example illustrates an MACA[25, 26] based two class pattern classifier.

### 4.1. PSMACA Tree Building

Input: Training set $S = \{S1, S2, \ldots, SK\}$
Output: PSMACA Tree.
Partition $(S, K)$
*Step* 1: Generate a PSMACA with $k$ number of attractor basins.
*Step* 2: Distribute $S$ into $k$ attractor basins (nodes).
*Step* 3: Evaluate the distribution of examples in each attractor basin.
*Step* 4: If all the examples ($S'$) of an attractor basin (node) belong to only one class, then label the attractor basin (leaf node) for that class.
*Step* 5: If examples ($S'$) of an attractor basin belong to $K'$ number of classes, then Partition $(S', K')$.
*Step* 6: Stop.

### 4.2. Random Generation of Initial Population

To form the initial population, it must be ensured that each solution randomly generated is a combination of an $n$-bit DS with $2m$ number of attractor basins (Classifier #1) and an $m$-bit DV (Classifier #2). The chromosomes are randomly synthesized according to the following steps.
1. Randomly partition $n$ into $m$ number of integers such that $n1 + n2 + \cdots + nm = n$.
2. For each $ni$, randomly generate a valid Dependency Vector (DV).
3. Synthesize Dependency String (DS) through concatenation of $m$ number of DVs for Classifier #1.
4. Randomly synthesize an $m$-bit Dependency Vector (DV) for Classifier #2.
5. Synthesize a chromosome through concatenation of Classifier #1 and Classifier #2.

## 5. EXPERIMENTAL SETUP

• Select the target CA protein (amino acid sequence) $T$, whose structure is to be predicted.
• Perform a PSMACA search, using the primary amino acid sequence $Tp$ of the target CA protein $T$. The objective is being to locate a set of CA proteins, $S = \{S_1, S_2 \ldots\}$ of similar sequence.
• Select from $S$ the primary structure Bp of a base CA protein, with a significant match to the target CA protein. A PSMACA,[16, 18] search produces a measure of similarity between each CA protein in $S$ and the target CA protein $T$. Therefore, Bp can be chosen as the CA protein with the highest such value.
• Obtain the base CA protein's structure, Bs, from the PDB.
• Using Bp, create an input sequences Ib (corresponding to the base CA protein) by replacing each amino acid in the primary structure with its hydrophobia city value. The output sequences Ob is created by replacing the structural elements in Bs with the values, 200, 600, 800 for helix $C$, strand and coil respectively.
• Solve the system identification problem, by performing CA de convolution with the output sequences Ob and the input sequence Ib to obtain the CA response, or the sought after running the algorithm.
• Transform the amino acid sequence of $Tp$ into a discrete time sequences It, and convolve with $F$; thereby producing the predicted structure ($Ot = It * F$) of the target CA protein.
• The result of this calculation $Ot$ is a vector of numerical values. For values between 0 and 200, a helix $C$ is predicted, and between 600 and 800, a strand is predicted by CA. All other values will be predicted as a coil by MACA. This produces mapping for the required target structure $Ts$ of the target CA protein $T$.

### 5.1. Sample Output

```
Amino Acids:
  MFRTKRSALVRRLWRSRAPGGEDEEEGAGGGGGGELR
GEGATDSRAHGAGGGGPGRAGCCLGKAVRGAKGHHHPHPP
AAGAGAAGGAEADLKALTHSVLKKLKERQLELLLQAVESR
GGTRTACLLLPGRLDCRLGPGAPAGAQPAQPPSSYSLPLL
LCKVFRWPDLRHSSEVKRLCCCESYGKINPELVCCNPHHL
SRLCELESPPPPYSRYPMDFLKPTADCPDAVPSSAETGGT
NYLAPGGLSDSQLLLEPGDRSHWCVVAYWEEKTRVGRLYC
VQEPSLDIFYDLPQGNGFCLGQLNSDNKSQLVQKVRSKIG
CGIQLTREVDGVWVYNRSSYPIFIKSATLDNPDSRTLLVH
KVFPGFSIKAFDYEKAYSLQRPNDHEFMQQPWTGFTVQIS
FVKGWGQCYTRQFISSCPCWLEVIFNSR

Predicted Structure:
  CCCCCHHHHHHHHHHCCCCCCCCCCCCCCCCCCCCC
CCCCCCCCCCCCCCCCCCCCCCCCCCCCCCCCCCCCC
CCCCCCCCCHHHHHHHHHHHHHHHHHHHHHHHHHHHC
CCCCCCEEEECCCCCCCCCCCCCCCCCCCCCCCCCEE
EEEEECCCCCCCCCCCCCCCCCCCCCEEECCCCCH
EEECCCCCCCCCCCCCCCCCCCCCCCCCCCCCCCC
CCCCCCCCCCCCCCCCCCCEEEEEEEECCCECEEEE
CCCCCECCCCCCCCCEEEECCCCCHHHHHHHHHHHC
CEEEEEECCCEEEEECCCCCEEEECCCCCCCCCCEEE
ECCCCCEEEECHHHHHHCCCCCCCECCCCCCCEEEEEE
```

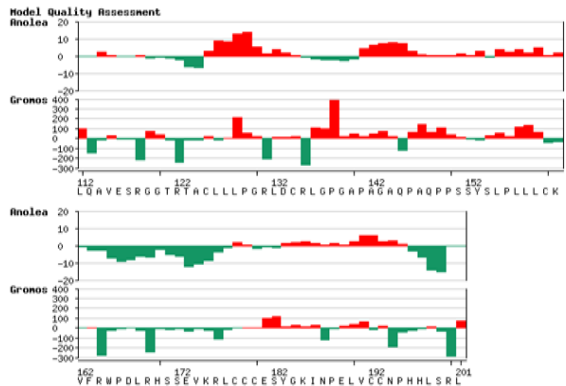

**Fig. 3.** Quality assessment graph.

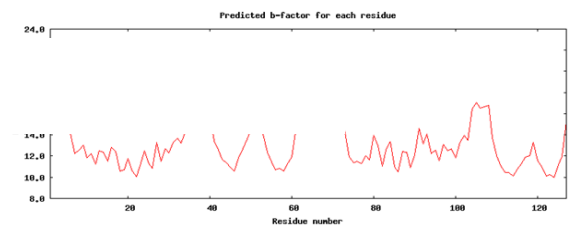

**Fig. 4.** *B* Factor residue.

```
EECCCCCCCCCCCCCCCCEEEEEEECCC

Predicted Solubility upon Over
expression:
SOLUBLE with probability 0.940939
```

## 6. EXPERIMENTAL RESULTS

In the experiments conducted, the base proteins are assigned the values 200,600,800 for helix *C*, strand and coil respectively. We have found an structure numbering scheme that is build on Boolean characters of CA which predicts the coils, stands and helices separately.

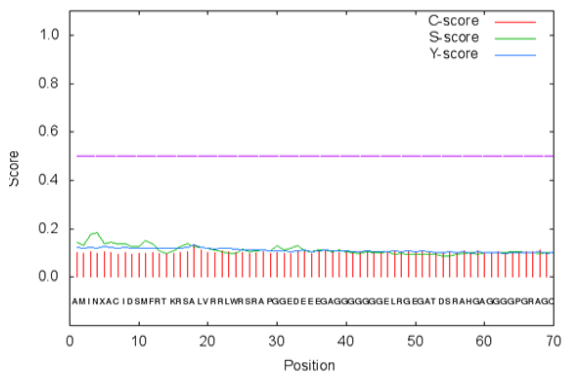

**Fig. 5.** *B* score and position of the sequence.

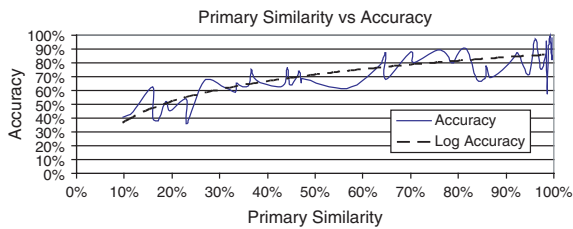

**Fig. 6.** Similarity accuracy.

| Target: 1PFC | Prediction accuracy (%) | Target: 1PP2 | Prediction accuracy (%) | Target: 1QL8 | Prediction accuracy (%) |
|---|---|---|---|---|---|
| Exp 1 | 62 | Exp 5 | 80 | Exp 9 | 82 |
| Exp 2 | 61 | Exp 6 | 90 | Exp 10 | 94 |
| Exp 3 | 65 | Exp 7 | 82 | Exp 11 | 83 |
| Exp 4 | 72 | Exp 8 | 85 | Exp 12 | 90 |

**Fig. 7.** Prediction Accuracy.

| Prediction method | Prediction accuracy for 1PFC (%) | Prediction accuracy for 1PP2 (%) | Prediction accuracy for 1QL8 (%) |
|---|---|---|---|
| DSP | 92 | 70 | 96 |
| PHD | 70 | 68 | 84 |
| SAM-T99 | 68 | 77 | 87 |
| SS Pro | 70 | 73 | 81 |
| PSMACA | 90 | 85 | 97 |

**Fig. 8.** Prediction accuracy for PSMACA.

The MACA based prediction procedure as described in the previous section is then executed, and each occurrence of each sequences in the resulting output, is predicted as shown in Figure 3 is predicted. The query sequence analyzer was designed and identification of the green terminals of the protein is simulated in the Figure 4. The analysis of the sequence and the place of joining of the proteins are also pointed out in the Figure 5. Experimental results Figures 6–8 which include the similarity and accuracy graph with each of the components are separately plotted.

## 7. CONCLUSION

To provide a more thorough analysis of the viability of our proposed technique more experiments will be conducted. Existing structure-prediction methods can predict the structure with 75% accuracy. Our preliminary results indicate that such a level of accuracy is attainable, and can be potentially surpassed with our method. PSMACA provides the best overall accuracy that ranges between 77% and 88.7% depending on the dataset.

## References and Notes

1. Debasis Mitra, M. Smith, Digital sequences processing in protein structure prediction. *Innovations in Applied Artificial Intelligence Lecture Notes in Computer Science* 3029, 40 (**2004**).
2. Sawcer, Stephen et al., Genetic risk and a primary role for cell-mediated immune mechanisms in multiple sclerosis. *Nature* 476, 214 (**2011**).
3. R. Abagyan, S. Batalov, T. Cardozo, M. Totrov, J. Webber, and Y. Zhou, Homology modeling with internal coordinate mechanics: Deformation zone mapping and improvements of models via conformational search. *Proteins*: *Structure, Function and Genetics* 1, 29 (**1997**).
4. N. Alexandrov and V. Solovyev, Effect of structure prediction on protein fold recognition and database search. *Genome Informatics* 7, 119 (**1996**).
5. C. B. Anfinsen, Principles that govern the folding of protein chains. *Science* 181, 223 (**1973**).
6. P. Baldi, S. Brunak, P. Frasconi, G. Pollastri, and G. Soda, Bidirectional dynamics for protein structure prediction, Sequence Learning: Paradigms, Algorithms and Applications, Springer (**2000**), pp. 80–104.
7. B. Boeckmann, A. Bairoch, R. Apweiler, M.-C. Blatter, A. Estreicher, E. Gasteiger, M. J. Martin, K. Michoud, C. O'Donovan, I. Phan, S. Pilbout, and M. Schneider, The SWISS-PROT protein knowledgebase and its supplement TrEMBL in 2003. *Nucleic Acids Res.* 31, 365 (**2003**).
8. R. Bonneau, J. Tsai, I. Ruczinski, D. Chivian, C. Rohl, C. Strauss, and D. Baker, Rosetta in CASP4: Progress in *ab initio* protein structure prediction. *PROTEINS*: *Structure, Function and Genetics* 5, 119 (**2001**).
9. E. P. Bourne and H. Weissig, Structural Bioinformatics, John Wiley & Sons (**2003**).
10. C. Brandon and J. Tooze, Introduction to Protein Structure, Garland Publishing (**1999**).
11. J. Chandonia and M. Karplus, New methods for accurate prediction of protein structure. *PROTEINS*: *Structure, Function and Genetics* 35, 293 (**1999**).
12. P. Chou and G. Fasman, Prediction of the structure of proteins from their amino acid sequence. *Advanced Enzymology* 47, 45 (**1978**).
13. A. Irback and E. Sandelin, On hydrophobicity correlations in protein chains. *Biophysical Journal* 79, 2252 (**2000**).
14. A. Irback, C. Peterson, and F. Potthast, Evidence for nonrandom hydrophobicity structures in protein chains. *Proc. Natl. Acad. Sci.* 93, 9533 (**1996**).
15. W. Kabsch and C. Sander, Dictionary of protein structure: Pattern recognition of hydrogen-bonded and geometrical features. *Biopolymers* 3, 2577 (**1983**).
16. K. Karplus, C. Barrett, and R. Hughey, Hidden markov models for detecting remote protein homologies. *Bioinformatics* 14, 846 (**1998**).
17. J. Skolnick and A. Kolinski, Computational studies of protein folding. *Computing in Science and Engineering* 3, 40 (**2001**).
18. R. Thiele, R. Zimmer, and T. Lengauer, Protein threading by recursive dynamic programming. *J. Mol. Biol.* 290, 757 (**1999**).
19. V. Veljkovic, I. Cosic, B. Dimitrijevic, and D. Lalovic, Is it possible to analyze DNA and protein sequences by the methods of digital sequences processing. *IEEE Transactions on Biomedical Engineering* BME-32, 337 (**1985**).
20. P. Kiran Sree and I. Ramesh Babu, Identification of protein coding regions in genomic DNA using unsupervised FMACA based pattern classifier. *International Journal of Computer Science and Network Security with ISSN*: 1738–7906 8, 1 (**2008**).
21. E. E. Snyder and G. D. Stormo, Identification of protein coding regions in genomic DNA. *ICCS Transactions* 248, 1 (**2002**).
22. E. E. Snyder and G. D. Stormo, Identification of coding regions in genomic DNA sequences: an application of dynamic programming and neural networks. *Nucleic Acids Res.* 11, 607 (**1993**).
23. P. Maji and P. P. Chaudhuri, FMACA: A fuzzy cellular automata based pattern classifier, *Proceedings of 9th International Conference on Database Systems*, Korea (**2004**), pp. 494–505.
24. P. Kiran Sree, Dr. Inampudi Ramesh Babu, and N. Usha Devi, Investigating an artificial immune system to strengthen the protein structure prediction and protein coding region identification using cellular automata classifier. *International Journal of Bioinformatics Research and Applications* 5, 647 (**2009**).
25. P. Kiran Sree, Dr. Inampudi Ramesh Babu, J. V. R. Murthy, P. Srinivasa Rao, and N. Usha Devi, Power-aware hybrid intrusion detection system (PHIDS) using cellular automata in wireless Ad Hoc networks. *World Scientific and Engineering Academy and Society TRANSACTIONS on COMPUTERS, USA,* 7, 1848 (**2008**).
26. P. Kiran Sree, Dr. Inampudi Ramesh Babu, et al., Identification of promoter region in genomic DNA using cellular automata based text clustering. *The International Arab Journal of Information Technology* (*IAJIT*) 7, 75 (**2010**).
27. P. Kiran Sree, Dr. Inampudi Ramesh Babu, et al., Improving quality of clustering using cellular automata for information retrieval. *International Journal of Computer Science,* (*Science Publications-USA*) 4, 167 (**2008**).
28. P. Kiran Sree, Dr. Inampudi Ramesh Babu, et al., A novel protein coding region identifying tool using cellular automata classifier with trust-region method and parallel scan algorithm (NPCRITCACA), *International Journal of Biotechnology and Biochemistry* (*IJBB*) 4, 177 (**2008**).
29. P. Kiran Sree, Dr. Inampudi Ramesh Babu, and N. Usha Devi, Non linear cellular automata in identification of protein coding regions. *Journal of Proteomics and Bioinformatics* (*USA*) 5, 123 (**2012**).